# Containment Graphs, Posets, and Related Classes of Graphs

MARTIN CHARLES GOLUMBIC[a] AND
EDWARD R. SCHEINERMAN[b]

[a]I.B.M. Israel Scientific Center
Technion City, Haifa 32000, Israel

[b]Department of Mathematical Sciences
The Johns Hopkins University
Baltimore, Maryland 21218

## INTRODUCTION

In this paper we introduce the notion of the containment graph of a family of sets and containment classes of graphs and posets. Let $\Sigma$ be a family of nonempty sets. We call a (simple, finite) graph $G = (V, E)$ a $\Sigma$-*containment graph* provided one can assign to each vertex $v_i \in V$ a set $S_i \in \Sigma$ such that $v_i v_j \in E$ if and only if $S_i \subset S_j$ or $S_i \supset S_j$. Similarly, we call a (strict) partially ordered set $P = (V, <)$ a $\Sigma$-*containment poset* if to each $v_i \in V$ we can assign a set $S_i \in \Sigma$ such that $v_i < v_j$ if and only if $S_i \subset S_j$. Obviously, $G$ is the comparability graph of $P$. The function $f\colon V \to \Sigma$, which assigns sets of $\Sigma$ to elements of $V$ by $f(v_i) = S_i$, is called a $\Sigma$-*containment representation* for $G$ and $P$, or simply a $\Sigma$-*representation*.

There are two approaches that one might take in investigating containment graphs. In the first, one restricts the family of sets $\Sigma$ to a certain type (such as intervals on a line, circles in the plane, paths in a tree) and then asks what class of graphs is obtained. In the second approach, one begins with a class **C** of comparability graphs and asks whether **C** can be characterized as the family of containment graphs of some family of sets.

In the next section we give some basic results on containment graphs and investigate the containment graphs of iso-oriented boxes in $d$-space. In the section following we present a characterization of those classes of posets and graphs that have containment representations by sets of a specific type. In the next section we present an alternative characterization, and then in the following section we extend our results to "injective" containment classes. After that we discuss similar characterizations for intersection, overlap, and disjointedness classes of graphs. Finally, in the last section we discuss the nonexistence of a characterization theorem for "strong" containment classes of graphs.

Unless specifically stated otherwise, all graphs and posets are assumed to be finite.

## CONTAINMENT GRAPHS AND POSETS

A binary relation $<$ on a set $V$ is called a *strict partial order* if it is irreflexive and transitive. The *comparability graph* of a partially ordered set (poset) $P = (V, <)$

192

is the undirected graph $G = (V, E)$ where $xy \in E$ if and only if $x$ and $y$ are comparable in $P$ (i.e., either $x < y$ or $y < x$). An undirected graph $G$ is a *comparability graph* if it is the comparability graph of some poset. Equivalently, $G$ is a comparability graph (or *transitively orientable graph*) if there exists an orientation $F$ of its edges such that for all $x, y, z \in V$,

$$xy \in F, \quad yz \in F \qquad \text{implies} \qquad xz \in F.$$

Such a transitive orientation $F$ is a strict partial order.

THEOREM 1: The following conditions are equivalent:

(i)   $G$ is a containment graph,
(ii)  $G$ is a comparability graph,
(iii) $G$ is a containment graph of subtrees of a tree.

*Proof:* (i) $\Rightarrow$ (ii). Let $G$ be the containment graph with $v_i$ assigned to a set $S_i$. Define an orientation $F$ of $G$ by $v_i v_j \in F$ iff $S_i \subset S_j$. Clearly, $F$ is transitive, so $G$ is a comparability graph.

(ii) $\Rightarrow$ (iii). Let $F$ be a transitive orientation of $G$. Let $T$ be a star with central vertex $x$ and pendant vertices numbered $1, \ldots, n$. To each vertex $v_j$ we associate the subtree $T_j$ consisting of the central vertex $x$ and those pendant vertices $i$ such that $v_i v_j \in F$. Observe that $T_i \subset T_j$ iff $v_i v_j \in F$. Hence the assignment of $T_i$ to $v_i$ gives a containment representation of $G$.

(iii) $\Rightarrow$ (i). Trivial.   □

Next, we investigate the class of containment graphs of rectilinear boxes (with sides parallel to the axes) in $d$-dimensional Euclidean space. These graphs are also called *box containment graphs* in $d$-space. The special case $d = 1$, the *interval containment graphs*, has appeared in the literature and is characterized by the following result:

THEOREM 2 [3]: The following conditions are equivalent:

(i)   A graph $G$ is the containment graph of intervals on line,
(ii)  $G$ is the comparability graph of a poset of dimension at most 2,
(iii) $G$ and its complement, $\bar{G}$, are both comparability graphs.

REMARK 1: The graphs that satisfy the preceding theorem are equivalent to the class of *permutation graphs* [4, 13]. They satisfy: The complement of a permutation graph is also a permutation graph.

We prove the analog of Theorem 2 for box containment graphs in $d$-dimensional Euclidean space.

A *linear order* is a poset in which any two elements are comparable. Let $P = (V, <)$ be a poset. A *realizer* of $P$ of size $k$ is a collection of linear orders $L_1 = (V, <_1), \ldots, L_k = (V, <_k)$ such that

$$x < y \quad \text{if and only if} \quad x <_i y \qquad \text{for all } i$$

that is, $P = L_1 \cap \cdots \cap L_k$. Every poset can be obtained as the intersection of some number of linear orders.

Dushnik and Miller [3] define the *dimension* of a poset $P$, denoted dim $P$, to be the size of the smallest possible realizer of $P$. Such a realizer is called a *minimum realizer* of $P$. Ore [12] observed the following equivalent definition of dimension. Let $R^k$ denote the poset of all $k$-tuples of real numbers ordered by $(x_1, \ldots, x_k) \leq (y_1, \ldots, y_k)$ if and only if each $x_i \leq y_i$. We say that $P$ can be *embedded* in $k$-space if it is isomorphic to a subposet of $R^k$. Then the dimension of $P$ is equal to the smallest $k$ for which $P$ can be embedded in $k$-space. The partial order dimension has been shown to be an invariant of the comparability graph:

THEOREM 3 [16]: If two partial orders $P$ and $P'$ have the same comparability graph, then dim $P = $ dim $P'$.

Therefore, for a comparability graph $G$ we define its *partial order dimension* to be the common dimension of all transitive orientations of $G$.

Let $G_1 = (V, E_1), \ldots, G_k = (V, E_k)$ be graphs with the same vertex set. Their *intersection* $G = G_1 \cap \cdots \cap G_k$ is defined to be the graph $G = (V, E_1 \cap \cdots \cap E_k)$. Their *union* $G' = G_1 \cup \cdots \cup G_k$ is defined to be the graph $G' = (V, E_1 \cup \cdots \cup E_k)$. By DeMorgan's law we have

$$G = G_1 \cap \cdots \cap G_k \qquad \text{iff} \qquad \bar{G} = \overline{G_1} \cup \cdots \cup \overline{G_k}.$$

THEOREM 4: Let $P$ be a poset. The following conditions are equivalent:

(i)  $P$ is the containment poset of boxes in $d$-space,
(ii)  The partial order dimension of $P$ is at most $2d$,
(iii)  $P$ is the intersection of $d$ interval containment posets.

*Proof:* (i)⇔(ii). Let $\{B_1, \ldots, B_n\}$ be a box containment representation of $P$ in $d$-space. Let $[a_{ik}, b_{ik}]$ be the interval obtained by projecting $B_i$ down to the $k$th axis, and choose $m > b_{ik}$ for all $i$ and $k$. We now map $B_i$ to the point $p_i = (a_{i1}, m - b_{i1}, \ldots, a_{id}, m - b_{id})$ in $R^{2d}$. Clearly, $B_i \subset B_j$ iff $p_i < p_j$ in the ordering of $R^{2d}$, which proves (ii).

The converse of this construction also works. Suppose that $P$ is embedded in $R^{2d}$ and let $m > c_{ik}$ for every point $p_i = (a_{i1}, c_{i1}, \ldots, a_{id}, c_{id})$ in the embedding and every $k$. Put $I_{ik} = [a_{ik}, m - c_{ik}]$ for $1 \leq k \leq d$. Let $B_i$ be the Cartesian product $I_{i1} \times \cdots \times I_{id}$, which is a box in $d$-space. Clearly, $B_i \subset B_j$ iff $p_i < p_j$. Thus the set $\{B_i\}$ is a box representation of $P$, proving (i).

(ii)⇔(iii). As stated in Theorem 2, the interval containment posets are precisely the posets of dimension at most 2. Clearly, $P$ is the intersection of $d$ posets of dimension 2 iff $P$ is the intersection of $2d$ linear orders. This concludes the proof of Theorem 4. □

THEOREM 5: Let $G$ be a comparability graph. The following conditions are equivalent:

(i)  $G$ is the containment graph of boxes in $d$-space,
(ii)  The partial order dimension of $G$ is at most $2d$,
(iii)  $G$ is the intersection of $2d - 1$ interval containment graphs,
(iv)  Every transitive orientation of $G$ is a containment poset of boxes in $d$-space.

*Proof:* (i) ⟹ (ii). If $G$ is a containment graph of boxes in $d$-space, then $G$ has a transitive orientation $P$ that is a containment poset of boxes in $d$-space. By Theorem 4, dim $P \leq 2d$. Hence by Theorem 3, dim $(G) \leq 2d$.

(ii) ⟹ (iv). Assume that dim $(G) \leq 2d$ and let $P$ be any transitive orientation of $G$. By Theorem 4, dim $(P) \leq 2d$, so Theorem 4 implies that $P$ is a containment poset of boxes in $d$-space.

(iv) ⟹ (i). Trivial.

(ii) ⟺ (iii). Golumbic *et al.* [10] proved that dim $(G) \leq k$ iff $\bar{G}$ is the union of $k - 1$ permutation graphs. By DeMorgan's law and Remark 1, this holds iff $G$ is the intersection of $k - 1$ interval containment graphs.    □

COROLLARY 1: The problem of deciding whether an arbitrary comparability graph is the containment graph of boxes in $d$-space is NP-complete for any fixed $d \geq 2$.

*Proof:* The problem of deciding whether the partial order dimension is at most $k$ is known to be NP-complete for any $k \geq 3$ [17]. Thus, it is NP-complete to recognize any of the statements of the theorem for any integer $d \geq 2$.    □    (For $d = 1$, the problem is polynomial.)

COROLLARY 2: Every comparability graph on $n$ vertices is the containment graph of boxes in $\lceil n/4 \rceil$-dimensional Euclidean space.

*Proof:* By Hiraguchi's theorem [11], dim $(G) \leq \lceil n/2 \rceil$. Therefore, by Theorem 5, the "box containment dimension" of $G$ is at most $\lceil n/4 \rceil$.    □

REMARK 2: W. T. Trotter (personal communication) has observed that the containment graphs of similar iso-oriented triangles in the plane are exactly the comparability graphs of partial order dimension at most 3.

## CONTAINMENT CLASSES

In this section we characterize those classes of graphs and posets that arise as containment classes. Our results are strikingly similar to the characterization of intersection classes in [15].

Let $\Sigma$ be a set of nonempty subsets, and let $\mathbf{P}(\Sigma)$ denote the class of all $\Sigma$-containment posets, that is, all posets that admit a $\Sigma$-containment representation. Similarly, let $\mathbf{G}(\Sigma)$ denote the class of all $\Sigma$-containment graphs.

We call $\mathbf{P}$ a *containment class* of posets if $\mathbf{P} = \mathbf{P}(\Sigma)$ for some $\Sigma$, and call $\mathbf{G}$ a *containment class* of graphs if $\mathbf{G} = \mathbf{G}(\Sigma)$ for some $\Sigma$. Finally, we say that $\mathbf{G}(\Sigma)$ is a *strong* containment class if for *every* transitive orientation $F$ of $G \in \mathbf{G}(\Sigma)$ we have $F \in \mathbf{P}(\Sigma)$.

We know from Theorem 1 that the class of all comparability (transitively orientable) graphs is the largest containment class of graphs. Furthermore, it is easy to see that $G$ is a $\Sigma$-containment graph if and only if $G$ has a transitive orientation that is a $\Sigma$-containment poset, that is, $\mathbf{G}(\Sigma) = \text{comp } (\mathbf{P}(\Sigma))$, where comp $(\mathbf{P})$ denotes the class of all comparability graphs of posets from $\mathbf{P}$.

A question of primary interest to us is the following: Given a class **G** of graphs, when does there exist a representation family $\Sigma$ such that $\mathbf{G} = \mathbf{G}(\Sigma)$?

EXAMPLE 1: The comparability graphs of lattices are not a containment class. To see this we note that a containment class must be hereditary (closed under induced subgraphs) and the lattice of subsets $\emptyset, \{a\}, \{b\}, \{a, b\}$ ordered by inclusion has an induced subposet $\emptyset, \{a\}, \{b\}$, which is not a lattice.

EXAMPLE 2 [Corneil and Golumbic, unpublished]: The containment graphs of paths in a tree is a containment class (by definition), but is *not* a strong class. In FIGURE 1 we give a representation of the transitive orientation $F$ of the 8-wheel, but we claim that the transitive orientation $F'$ has no representation. (*Note:* In the figure directed edges $x \rightarrow y$ are to be interpreted as $x > y$ in the corresponding poset.) For if the central vertex corresponds to a path that contains the remaining eight paths, then we would have an interval containment representation for the chordless 8-cycle, which is not possible.

EXAMPLE 3: As shown in Theorem 5, the containment graphs of boxes in $d$-space form a strong containment class. The following are also strong containment classes: The comparability graphs of interval orders, of weak orders, and of semiorders (see [7, sec. 6]).

A poset, $P' = (X', <')$ is called an *induced subposet* of $P = (X, <)$ if $X' \subseteq X$ and for $x, y \in X'$, $x <' y$ iff $x < y$. Our notation is $P' \leq P$. A class **P** of posets is called *hereditary* provided $P \in \mathbf{P}$ and $P' \leq P$ imply $P' \in \mathbf{P}$. A *composition sequence* for a class of posets, **P**, is a sequence of posets $P_1, P_2, \ldots$, such that

(1) each $P_i \in \mathbf{P}$,
(2) $P_i \leq P_{i+1}$ for all $i$, and
(3) if $P \in \mathbf{P}$, then $P \leq P_k$ for some $k$.

Given a poset $P = (X, <)$ we can define an equivalence relation $\approx$ on $X$ as follows: For $x, y \in X$, $x \approx y$, provided either (1) $x = y$, or (2) $x$ and $y$ are incomparable and for all other $z \in X$, $x < z$ iff $y < z$ and $z < x$ iff $z < y$. Next, we define $r(P)$ to be the poset $P$ reduced modulo $\approx$, that is, $r(P)$ is an induced subposet of $P$ formed by taking one element in each $\approx$ equivalence class. Since all such induced subposets are isomorphic, $r(P)$ is uniquely defined. We say that $P'$ is obtained from $P$ by *vertex multiplication* provided $P \leq P'$ and $r(P) = r(P')$. Finally, **P** is *closed* under vertex multiplication iff $r(P) \in \mathbf{P}$ implies that $P \in \mathbf{P}$.

THEOREM 6: Let **P** be a family of posets. **P** is a containment class if and only if **P** is hereditary, has a composition sequence, and is closed under vertex multiplication.

*Proof:* Suppose $\mathbf{P} = \mathbf{P}(\Sigma)$ and let $P \in \mathbf{P}$ have a $\Sigma$-representation function $f$. Obviously **P** is hereditary since any subposet $P' \leq P$ is also a $\Sigma$-containment poset by restricting the domain of $f$. If $P''$ is obtained from $P$ by vertex multiplication, then we simply extend $f$ to $P''$ by defining $f''(x_i) = f(x)$ for every $x_i$ in the equivalence class corresponding to $x$ in $P$. Clearly, $f''$ is a $\Sigma$-containment representation for $P''$, so **P** is closed under vertex multiplication. Finally, since **P** is countable, we may assume $\Sigma$ is countable as well. [One readily checks that for any $\Sigma$ there exists a countable family $\Sigma' \subseteq \Sigma$ with $\mathbf{P}(\Sigma') = \mathbf{P}(\Sigma)$.] Let $\Sigma = \{S_1, S_2, S_3, \ldots\}$. For all $k$ we define $P_k$ to be the containment poset of $k$ copies each of $S_1, \ldots, S_k$. Clearly, each $P_k$ is in $P(\Sigma)$ and $P_{k-1} \leq P_k$. For any $P = (X, <) \in \mathbf{P}$ we let

$$k = \max \{|X|, \max \{i : f(x) = S_i \text{ and } x \in X\}\}.$$

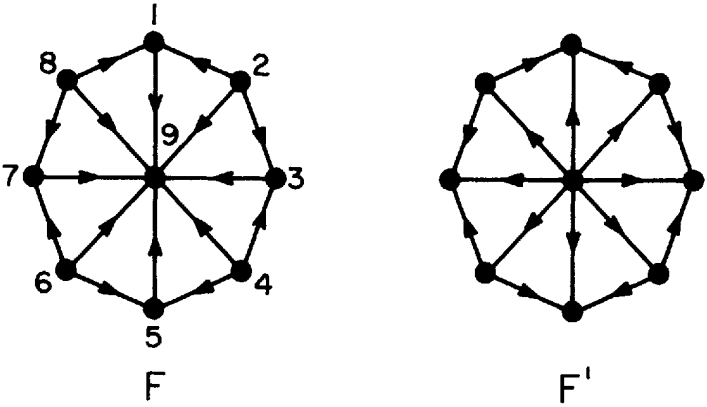

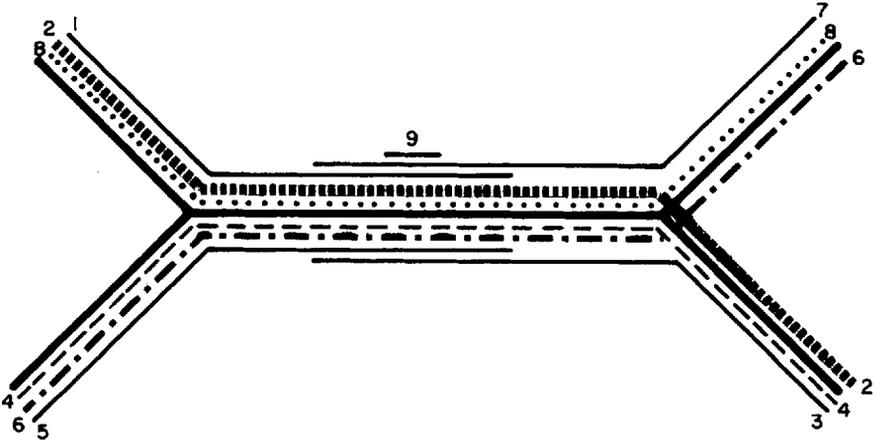

**FIGURE 1.**

It follows that $P \leq P_k$, so $P_1 \leq P_2 \leq \cdots$ forms a composition sequence for $\mathbf{P}$.

Conversely, suppose $\mathbf{P}$ satisfies the three conditions of the theorem. Let $\{P_i : i = 1, 2, \ldots\}$ be a composition sequence. Since $\mathbf{P}$ is hereditary, we may assume each $P_i = (X_i, <)$ has $|X_i| = i$ and $X_i \subset X_{i+1}$. Thus, $X_i = \{x_1, \ldots, x_i\}$. We now define sets $S_k$ to form our family $\Sigma$ by $S_k = \{i: x_i \leq x_k\}$. Let $\Sigma = \{S_1, S_2, \ldots\}$. Observe that $f: X_k \to \Sigma$ by $f(x_i) = S_i$ is a $\Sigma$-containment representation of $P_k$. If $P \in \mathbf{P}$, then $P \leq P_k \in \mathbf{P}(\Sigma)$ for some $k$. It follows that $P \in \mathbf{P}(\Sigma)$. Thus $\mathbf{P} \subseteq \mathbf{P}(\Sigma)$.

To show the opposite inclusion, suppose $P = (X, <) \in \mathbf{P}(\Sigma)$ and let $f: X \to \Sigma$ be a containment representation. Let $k_1$ be the largest subscript of an $S_i$ used in the representation $f$, and let $k_2$ be maximum number of times any set of $\Sigma$ is assigned to an element of $X$. Let $k = \max\{k_1, k_2\}$. Let $P'$ denote the poset formed by multiplying each element of $P_k$ $k$ times. Thus $P' \in \mathbf{P}$, since $\mathbf{P}$ is closed under vertex multiplication. One readily verifies $P \leq P'$, hence, $P \in \mathbf{P}$ by the hereditary property. Thus, $\mathbf{P}(\Sigma) \subseteq \mathbf{P}$, and the theorem follows. $\square$

A graph $G$ is an *induced subgraph* of $H$ provided $V(G) \subseteq V(H)$ and $E(G) = \{xy: x, y \in V(G)$ and $xy \in E(H)\}$. We write $G \leq H$ in this case. A class of graphs $\mathbf{G}$ is hereditary if $G \leq H \in \mathbf{G}$ implies $G \in \mathbf{G}$. A *composition sequence* for a class of graphs $\mathbf{G}$ is a sequence $G_1, G_2, G_3, \ldots$, with (1) each $G_i \in \mathbf{G}$, (2) $G_i \leq G_{i+1}$ for all $i$, and (3) if $G \in \mathbf{G}$, then $G \leq G_k$ for some $k$. A composition sequence is called *coherently transitively orientable*, provided each graph in the sequence can be transitively oriented so that $x \to y$ in $G_i$ implies $x \to y$ in $G_{i+1}$.

Vertex multiplication is also defined for graphs in a manner analogous to the previous definition for posets (see, for example, [8]). Let us write $v \sim w$ when vertices $v$ and $w$ are adjacent. Two vertices of a graph are equivalent, $v \approx w$ if either $v = w$ or else $v$ and $w$ are not adjacent, and for all other vertices $z$ we have $z \sim v$ iff $z \sim w$. The graph $\rho(G)$ is obtained by taking an induced subgraph of $G$ with one vertex per $\approx$ class. A class of graphs $\mathbf{G}$ is *closed under vertex multiplication* if $\rho(G) \in \mathbf{G}$ implies $G \in \mathbf{G}$.

As in the poset setting, the conditions together are necessary and sufficient.

THEOREM 7: Let $\mathbf{G}$ be a class of graphs. $\mathbf{G}$ is a containment class if and only if $\mathbf{G}$ is hereditary, has a coherently transitively orientable composition sequence, and is closed under vertex multiplication.

*Proof:* Suppose $\mathbf{G}$ is the containment class of $\Sigma$, that is, $\mathbf{G} = \mathbf{G}(\Sigma)$, and let $\mathbf{P} = \mathbf{P}(\Sigma)$. Obviously, $\mathbf{G}$ is hereditary. By Theorem 6, $\mathbf{P}$ has a composition sequence $P_1 \leq P_2 \leq \cdots$. Put $G_i = \text{comp}(P_i)$ for all $i$. It is immediate that $G_1 \leq G_2 \leq \cdots$ is a composition sequence for $\mathbf{G}$. Moreover, the natural orientations inherited from the $P_i$ form a coherent transitive orientation for the sequence. Finally, we must show that $\mathbf{G}$ is closed under vertex multiplication. Suppose $\rho(G) \in \mathbf{G}$ and let $f: V(\rho(G)) \to \Sigma$ be a $\Sigma$-representation. As in the proof of Theorem 6, we extend $f$ to $f''$ by defining $f''(v_i) = f(v)$ for every $v_i$ in the equivalence class corresponding to $v$ in $G$. Clearly $f''$ is a $\Sigma$-representation for $G$, hence $G \in \mathbf{G}$.

Conversely, suppose $\mathbf{G}$ is hereditary, has $G_1 \leq G_2 \leq \cdots$ as its composition sequence, and is closed under vertex multiplication. Orient each graph transitively so that the orientations are coherent. Each graph $G$ in $\mathbf{G}$ can be transitively oriented, since $G \leq G_k$ for some $k$, and we orient $G$ according to the orientation of $G_k$. Thus every graph in $\mathbf{G}$ has been given a transitive orientation.

Since a transitive orientation of a graph $G$ can be considered a poset $P$ with $x > y$ if and only if $x \to y$, the class $\mathbf{P}$ of all posets derived in this way from $\mathbf{G}$ is clearly hereditary with composition sequence $P_1 \leq P_2 \leq \cdots$, and is closed under vertex multiplication. Thus, by Theorem 6, **P** is a containment class of posets, **P** = **P**($\Sigma$). It follows easily that **G** = comp (**P**), and so **G** = **G**($\Sigma$).  □

## AN ALTERNATE CHARACTERIZATION

In the previous section we found necessary and sufficient conditions that a class of graphs must satisfy in order to be a containment class. In some respects this solves the problem at hand. However, the coherence part of the "coherently transitively orientable" condition is, in general, difficult to directly verify. Even if we know that all the graphs in our class are transitively orientable, it is not clear that we can give a coherent orientation to the graphs in the sequence. For example, if we orient $G_1$, and then attempt to extend that orientation to $G_2$, and then to $G_3$, etc., we are doomed to failure. Such a failed construction is depicted in FIGURE 2. One cannot be greedy when constructing a transitive orientation for a graph. In this section we present a more tractable characterization.

A *countable* graph is a graph whose vertex set we allow to be finite or countably infinite. Given a sequence of graphs $G_1 \le G_2 \le G_3 \cdots$, we define the limit of this sequence to be the countable graph whose vertex set is the union $\bigcup_{i \ge 1} V(G_i)$, in which $x \sim y$ if and only if they are adjacent in some $G_k$; one writes $G = \lim_{i \to \infty} G_i$. Note that every finite induced subgraph of the limit graph $G$ is an induced subgraph of some $G_k$.

It is clear that a transitive orientation for the limit graph implies a coherent transitive orientation for the individual graphs in the sequence. The converse, however, is also true:

LEMMA: A sequence of graphs $G_1 \le G_2 \le G_3 \le \cdots$ has a coherent transitive orientation if and only if $\lim_{t \to \infty} G_i$ has a transitive orientation.

*Proof:* This follows immediately from the characterization of comparability graphs of Gilmore and Hoffman [6]. Their characterization, which applies to finite as well as to infinite graphs, states:

A graph is a comparability graph if and only if each odd "cycle" has at least one triangular chord. [Here, a "cycle" may retrace over edges. A "triangular chord" is an edge of the form $v_i v_{i+2}$.]

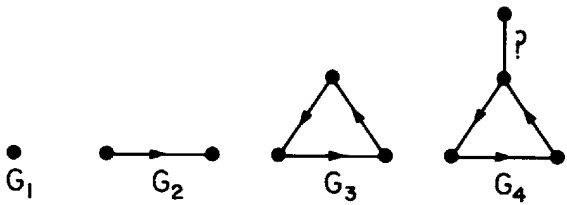

**FIGURE 2.**

One need only note that this condition is "locally finite" and a graph fails to be transitively orientable if and only if some finite induced subgraph is not transitively orientable. □

We can now state our alternate characterization:

THEOREM 8: A class of graphs is a containment class if and only if it is a hereditary class of transitively orientable graphs with a composition sequence, and is closed under vertex multiplication.

*Proof:* The necessity of these conditions is immediate. Suppose **G** is hereditary, closed under vertex multiplication, contains only transitively orientable graphs, and has $G_1 \leq G_2 \leq \cdots$ as a composition sequence. By Theorem 7, we need only show that the sequence is coherently transitively orientable. We know that each $G_i$ has a transitive orientation. Let $G = \lim_{i \to \infty} G_i$. Observe that $G$ must have a transitive orientation, for otherwise one of its finite induced subgraphs would fail to have a transitive orientation. Orient $G$ transitively. Now assign to each $G_i$ the orientation it inherits from $G$. Clearly, the sequence now has a coherent transitive orientation. □

## INJECTIVE CONTAINMENT REPRESENTATIONS

Thus far the containment representation functions we have considered, $f: V \to \Sigma$ were allowed to assign the same set to more than one element. This resulted in a collection of vertices all of which were equivalent. We now consider the effect of requiring representation functions $f$ to be one-to-one. Denote by $\mathbf{P}_1(\Sigma)$ [respectively, $\mathbf{G}_1(\Sigma)$] the class of all posets [graphs] with a containment representation that is one-to-one. Such a class is called an *injective* containment class of posets [graphs].

It is clear that injective containment classes necessarily are hereditary and possess composition sequences. Together, these conditions are sufficient:

THEOREM 9: A class of posets is an injective containment class if and only if it has a composition sequence and is hereditary.

*Proof:* Since the necessity of the conditions is immediate, we consider the sufficiency. Let **P** be a hereditary class of posets with a composition sequence $P_1 \leq P_2 \leq \cdots$. As in the proof of Theorem 6, we may assume that $P_i = (\{x_1, \ldots, x_i\}, <)$ and set $S_k = \{i: x_i \leq x_k\}$. Put $\Sigma = \{S_1, S_2, \ldots\}$. We claim that $\mathbf{P} = \mathbf{P}_1(\Sigma)$. Observe that $P_k$ is the containment poset of the sets $S_1, \ldots, S_k$. Hence, if $P \in \mathbf{P}_1(\Sigma)$, then $P \leq P_k$ for $k$ sufficiently large. Hence $P \in \mathbf{P}$, since **P** is hereditary. On the other hand, suppose $P \in \mathbf{P}$. By the definition of composition sequence $P \leq P_k$ for some $k$, and it follows that $P \in \mathbf{P}_1(\Sigma)$. □

COROLLARY 3: A class of graphs is an injective containment class if and only if it is a hereditary class of transitively orientable graphs with a composition sequence.

It follows that every containment class is an injective containment class, but the converse is not necessarily true. For example, if $\Sigma = \{A\}$, where $A$ is any nonempty set, then $\mathbf{G}(\Sigma)$ contains all edgeless graphs while $\mathbf{G}(\Sigma)$ contains only the trivial graph with one vertex.

## RELATED CLASSES OF GRAPHS: INTERSECTION, OVERLAP, AND DISJOINTEDNESS

The following related problem was discussed in [15]. A function $f: V(G) \to \Sigma$ is called a $\Sigma$-*intersection* representation of a graph $G$ if and only if for all vertices $v$, $w$ we have $v \sim w$ if and only if $f(v) \cap f(w) \neq \emptyset$. $G$ is called a $\Sigma$-*intersection graph* and the class of all such graphs is called an *intersection class* and is denoted $\Omega(\Sigma)$. If the representing functions are required to be injections, the resulting class is called an *injective* intersection class and is denoted $\Omega_1(\Sigma)$.

Analogous to the concept of vertex multiplication, we have the notion of vertex *expansion*. For $v$, $w \in V(G)$ we write $v \approx' w$ if either $v = w$ or else $v \sim w$, and for all other vertices $z$ we have $z \sim v$ iff $z \sim w$. Observe the $v \approx' w$ in $G$ if and only if $v \approx w$ in $\bar{G}$. We put $\rho'(G)$ to be the induced subgraph of $G$ formed by taking one vertex per $\approx'$ equivalence class. Finally, a class of graphs, **G**, is *closed under vertex expansion* if $\rho'(G) \in \mathbf{G}$ implies $G \in \mathbf{G}$.

THEOREM 10 [15]: A class of graphs is an intersection class if and only if it is hereditary, has a composition sequence, and is closed under vertex expansion. It is an injective intersection class if and only if it is hereditary and has a composition sequence.

It is natural when considering containment and intersection representations of graphs to consider two further representation schemes.

First, a graph is said to be a $\Sigma$-*overlap* graph, if it has a $\Sigma$-*overlap* representation: A function $f: V(G) \to \Sigma$ such that $v \sim w$ if and only if $f(v)$ *overlaps* $f(w)$, that is, $f(v) \cap f(\omega) \neq \emptyset$, but neither $f(v) \subseteq f(w)$ nor $f(v) \supseteq f(w)$. The class of all $\Sigma$-overlap graphs is denoted $\mathbf{O}(\Sigma)$. If the functions are restricted to injections, we arrive at the notion of *injective* overlap classes, denoted $\mathbf{O}_1(\Sigma)$.

Second, a graph is a $\Sigma$-*disjointedness* graph if it has a $\Sigma$-*disjointedness* representation: A function $f: V(G) \to \Sigma$ such that $v \sim w$ if and only if $f(v) \cap f(w) = \emptyset$. The family of all $\Sigma$-disjointedness graphs is $\Delta(\Sigma)$. If the representations are all injections, we arrive at $\Delta_1(\Sigma)$. We wish to characterize those classes of graphs that are [injective] overlap and disjointedness classes.

It is immediate that a graph $G$ is a $\Sigma$-disjointedness graph if and only if $\bar{G}$ is a $\Sigma$-intersection graph. Since $G \leq H$ if and only if $\bar{G} \leq \bar{H}$, and since $\rho(G) = \rho'(\bar{G})$, Theorem 10 gives:

THEOREM 11: A class of graphs is a disjointedness class if and only if it is hereditary, has a composition sequence, and is closed under vertex multiplication. It is an injective disjointedness class if and only if it is hereditary and has a composition sequence.

Next we consider overlap classes. One readily verifies that an [injective] overlap class is hereditary and has a composition sequence. Noninjective overlap classes are closed under vertex multiplication. We claim that these conditions are sufficient:

THEOREM 12: A class of graphs is an overlap class if and only if it is hereditary, has a composition sequence, and is closed under vertex multiplication. It is an injective intersection class if and only if it is hereditary and has a composition sequence.

*Proof:* We only need to show the sufficiency. We begin with the injective case: Let G be a class of graphs that is hereditary and has a composition sequence. It follows from Theorem 10 that G is an injective intersection class. Let $\Sigma$ be a family of sets with $G = \Omega_1(\Sigma)$. Without loss of generality we may assume $\Sigma$ is countable: $\Sigma = \{S_1, S_2, \ldots\}$. Let $X = \{x_1, x_2, \ldots\}$ be a set of elements such that no $x_i$ is in any $S_j$ for all $i$ and $j$. Let $T_i = S_i \cup \{x_i\}$ and $\Sigma' = \{T_1, T_2, \ldots\}$. It is now clear that $T_i$ overlaps $T_j$ if and only if $S_i$ intersects $S_j$. Thus $G = \Omega_1(\Sigma) = O_1(\Sigma')$.

Now suppose G is also closed under vertex multiplication. We know that $G = O_1(\Sigma') \subseteq O(\Sigma')$. On the other hand, suppose $G \in O(\Sigma')$, then $\rho(G) \in O(\Sigma')$ since $\rho(G) \leq G$. Now any $\Sigma'$-overlap representation of $\rho(G)$ is necessarily an injection; this is because no two vertices of $\rho(G)$ are $\approx$ equivalent, but if two vertices of $G$ were assigned to the same set of $\Sigma'$, they would be $\approx$ equivalent. Hence, $\rho(G) \in O_1(G) = G$. Since G is closed under vertex multiplication, $G \in G$, thus $G \supseteq O(\Sigma')$.   □

We can summarize our results as follows. The properties of classes of graphs we have considered are:

(1) hereditary,
(2) has a composition sequence,
(3) contains only transitively orientable graphs,
(4) is closed under vertex multiplication, and
(5) is closed under vertex expansion.

The necessary and sufficient conditions for the classes we have discussed are:

| Class | Injective | Noninjective |
|---|---|---|
| Containment | 1, 2, 3 | 1, 2, 3, 4 |
| Intersection | 1, 2 | 1, 2, 5 |
| Overlap | 1, 2 | 1, 2, 4 |
| Disjointedness | 1, 2 | 1, 2, 4 |

A class of graphs that satisfies all five properties would be simultaneously a containment, intersection, overlap, and disjointedness class of both the injective and noninjective varieties. For example, the family of all transitively orientable graphs can be represented in all these ways. Let $\Sigma_1$ denote family of subtrees of a tree and let $\Sigma_2$ denote the family of all Cartesian graphs of continuous real-value functions (i.e., the planar curves that are also referred to as "graphs"). Using [2, 5, 10] and our Theorem 1 we have:

$$G(\Sigma_1) = G_1(\Sigma_1) = O(\Sigma_1) = O_1(\Sigma_1) = \Omega(\Sigma_1) = \Omega_1(\Sigma_1) = \Delta(\Sigma_2) = \Delta_1(\Sigma_2)$$

all equal the class of transitively orientable graphs!

## ON CHARACTERIZING STRONG CONTAINMENT CLASSES

In the section on containment classes we defined a notion of *strong* containment class as follows: A containment class, $G(\Sigma)$, is *strong* if for every transitive orienta-tion $F$ of $G \in G(\Sigma)$ we have $F \in P(\Sigma)$. Example 3 shows that the containment graphs of real intervals form a strong containment class, while the containment class of edge paths in trees (Example 2) does not form a strong class.

It would be very desirable to have a characterization of strong containment classes similar to our characterization of containment classes. In this section we show that no such characterization is possible. The notion of a containment class of graphs is an *intrinsic* one: One need only examine the class of graphs $G$ to determine if it is a containment class. However, the notion of a strong containment class is an *extrinsic* one; it depends on the sets in $\Sigma$. One cannot say whether or not a containment class $G$ is strong or not. We prove this by showing that there exists two families of sets, $\Sigma$ and $\Sigma'$, with $G(\Sigma) = G(\Sigma')$, where $G(\Sigma)$ is not a strong containment class, but $G(\Sigma')$ is a strong class.

Let $\Sigma$ be the family of all edge paths in trees. Let $G = G(\Sigma)$; we know that $G(\Sigma)$ is not a strong containment class. Note that $G$ is closed under disjoint union, that is, if $G$ and $H$ are in $G$, then so is their disjoint union. Let $P$ denote the class of posets formed by taking all possible transitive orientations of all graphs in $G$. Clearly, $P$ is hereditary and closed under disjoint union of posets. Let $P_i$ denote the disjoint union of all posets in $P$ with at most $i$ elements. It follows that $P_1 \leq P_2 \leq \cdots$ is a composition sequence for $P$. Furthermore, one can deduce from $G$'s closure under vertex multiplication that $P$ is also closed under vertex multiplication. Thus $P$ is a containment class of posets by Theorem 6. Therefore $P = P(\Sigma')$ for some family $\Sigma'$. One now notes that $G(\Sigma') = G(\Sigma) = G$, yet every transitive orientation of every graph in $G$ is in $P = P(\Sigma')$, thus $G(\Sigma')$ is a strong containment class.

Thus it is impossible to characterize classes of graphs $G$ as strong or not strong; it depends on how the class is represented. This leaves us with a more difficult problem: Characterize those family of sets $\Sigma$ for which $G(\Sigma)$ is (or is not) a strong containment class.